\documentclass[iop]{emulateapj}

\usepackage[caption=false]{subfig} 
\usepackage{rotating}
\usepackage[usenames,dvipsnames]{xcolor}
\usepackage{color}

\newcommand{\lya}{Ly$\alpha$}
\newcommand{\halpha}{H$\alpha$}
\newcommand{\hbeta}{H$\beta$}
\newcommand{\hi}{H{\sc i}}

\newcommand{\lfuv}{$L_\mathrm{FUV}$}
\newcommand{\lsun}{$L_\odot$}

\newcommand{\ergsec}{erg~s$^{-1}$}

\newcommand{\ergseccmaa}{erg~s$^{-1}$~cm$^{-2}$~\AA$^{-1}$}
\newcommand{\ergseccm}{erg~s$^{-1}$~cm$^{-2}$}

\newcommand{\ebv}{$E_{B-V}$}

\newcommand{\rpt}{$R_{\mathrm{P}20}$}
\newcommand{\rptfuv}{$R_{\mathrm{P}20}^{\mathrm{FUV}}$}
\newcommand{\rptlya}{$R_{\mathrm{P}20}^{\mathrm{Ly}\alpha}$}
\newcommand{\rptha}{$R_{\mathrm{P}20}^{\mathrm{H}\alpha}$}
\newcommand{\rext}{$\xi_{\mathrm{Ly}\alpha}$}

\newcommand{\nII}{[N{\sc ii}]}

\slugcomment{Accepted by ApJL}

\shorttitle{Extended \lya\ Emission in Nearby Galaxies}
\shortauthors{M. Hayes et al.}

\begin{document}

\title{The Lyman alpha Reference Sample: Extended Lyman alpha Halos Produced at Low Dust Content}

\author{Matthew Hayes\altaffilmark{2,3},
G\"oran \"Ostlin\altaffilmark{4}, 
Daniel Schaerer\altaffilmark{5,3}, 
Anne Verhamme\altaffilmark{5}, 
J. Miguel Mas-Hesse\altaffilmark{6}, 
Angela Adamo\altaffilmark{7}, 
Hakim Atek\altaffilmark{8}, 
John M. Cannon\altaffilmark{9}, 
Florent Duval\altaffilmark{4}, 
Lucia Guaita\altaffilmark{4}, 
E. Christian Herenz\altaffilmark{10},
Daniel Kunth\altaffilmark{11}, 
Peter Laursen\altaffilmark{12}, 
Jens Melinder\altaffilmark{4}, 
Ivana Orlitov{\'a}\altaffilmark{5,13}, 
H\'ector Ot\'i-Floranes\altaffilmark{14,15,6}, and 
Andreas Sandberg\altaffilmark{4}
}

\email{matthew.hayes@irap.omp.eu}
\email{}
\altaffiltext{1}{Based on observations made with the NASA/ESA Hubble 
Space Telescope, obtained at the Space Telescope Science Institute, 
which is operated by the Association of Universities for Research in 
Astronomy, Inc., under NASA contract NAS 5-26555. These observations are
associated with program \#12310.}
\altaffiltext{2}{Universit\'e de Toulouse; UPS-OMP; IRAP; Toulouse, France}
\altaffiltext{3}{CNRS; IRAP; 14, avenue Edouard Belin, F-31400 Toulouse, France}
\altaffiltext{4}{Department of Astronomy, Oskar Klein Centre, Stockholm University, AlbaNova University Centre, SE-106 91 Stockholm, Sweden}
\altaffiltext{5}{Geneva Observatory, University of Geneva, 51 Chemin des Maillettes, CH-1290 Versoix, Switzerland}
\altaffiltext{6}{Centro de Astrobiolog\'ia (CSIC--INTA), Departamento de Astrof\'isica, POB 78, 28691 Villanueva de la Ca\~nada, Spain.}
\altaffiltext{7}{Max Planck Institute for Astronomy, K\"onigstuhl 17, D-69117 Heidelberg, Germany.}
\altaffiltext{8}{Laboratoire d’Astrophysique, \'Ecole Polytechnique F\'ed\'erale de Lausanne (EPFL), Observatoire, CH-1290 Sauverny, Switzerland.}
\altaffiltext{9}{Department of Physics and Astronomy, Macalester College, 1600 Grand Avenue, Saint Paul, MN 55105, USA.}
\altaffiltext{10}{Leibniz-Institut f\"ur Astrophysik (AIP), An der Sternwarte 16, D-14482 Potsdam, Germany.}
\altaffiltext{11}{Institut d'Astrophysique de Paris, UMR 7095 CNRS \& UPMC, 98 bis Bd Arago, 75014 Paris, France.}
\altaffiltext{12}{Dark Cosmology Centre, Niels Bohr Institute, University of Copenhagen, Juliane Maries Vej 30, 2100 Copenhagen, Denmark.}
\altaffiltext{13}{Astronomical Institute of the Academy of Sciences, Bo\v cn\i\'\ II 1401/1a, CZ-141 31 Praha 4, Czech Republic.}
\altaffiltext{14}{Instituto de Astronom\'ia, Universidad Nacional Aut\'onoma de M\'exico, Apdo. Postal 106, Ensenada B. C. 22800 Mexico}
\altaffiltext{15}{Dpto. de F\'isica Moderna, Facultad de Ciencias, Universidad de Cantabria, 39005 Santander, Spain}

\begin{abstract}
We report on new imaging observations of the Lyman alpha emission line
(\lya), performed with the Hubble Space Telescope, that comprise the 
backbone of the 
\emph{Lyman alpha Reference Sample} (LARS). We present images of 14 
starburst galaxies at redshifts $0.028 < z < 0.18$ in 
continuum-subtracted \lya, \halpha, 
and the far ultraviolet continuum. We show that \lya\ is 
emitted on scales that systematically exceed those of 
the massive stellar population and recombination nebulae: as measured 
by the 
Petrosian 20 percent radius, \rpt, \lya\ radii are larger than 
those of \halpha\ by factors ranging from 1 to 3.6, with an average of 
2.4. The average ratio of \lya-to-FUV radii is 2.9. This suggests that
much of the \lya\ light is pushed to large radii by resonance 
scattering. Defining the \emph{Relative Petrosian Extension} of \lya\
compared to \halpha, \rext = \rptlya /\rptha, we find 
\rext\ to be uncorrelated with total \lya\ luminosity.
However \rext\ is strongly correlated 
with quantities that scale with dust content, in the sense that a low 
dust abundance is a necessary requirement (although not the only one) 
in order to spread \lya\ photons throughout the interstellar medium and
drive a large extended \lya\ halo. 
\end{abstract}


\section{Introduction}

The Lyman alpha emission line (\lya), emitted by the spontaneous 
de-excitation over the $n=2\rightarrow 1$ electronic transition in 
neutral hydrogen (\hi), is now an established observational probe of 
evolving galaxies in the high-$z$
Universe \citep{Cowie1998,Rhoads2000}. Exploitation of \lya\
has resulted in significant galaxy surveys 
\citep{Ouchi2008,Nilsson2009survey,Guaita2010,Adams2011}, the next 
generations of which will recover vast numbers of galaxies. 
However the 
\hi\ abundance in most galaxies, combined with the large \lya\ absorption
cross section of ground-state hydrogen, suggests 
that most \lya\ will be absorbed and re-scattered
by the same transition that created it. Thus most \lya\ photons
are thought to be subject to multiple scattering events as they 
encounter neutral gas, resulting in a complicated radiative transport 
\citep{Neufeld1990,Verhamme2006,Laursen2009b}. 

\begin{figure*}
\centering
\includegraphics[scale=0.42]{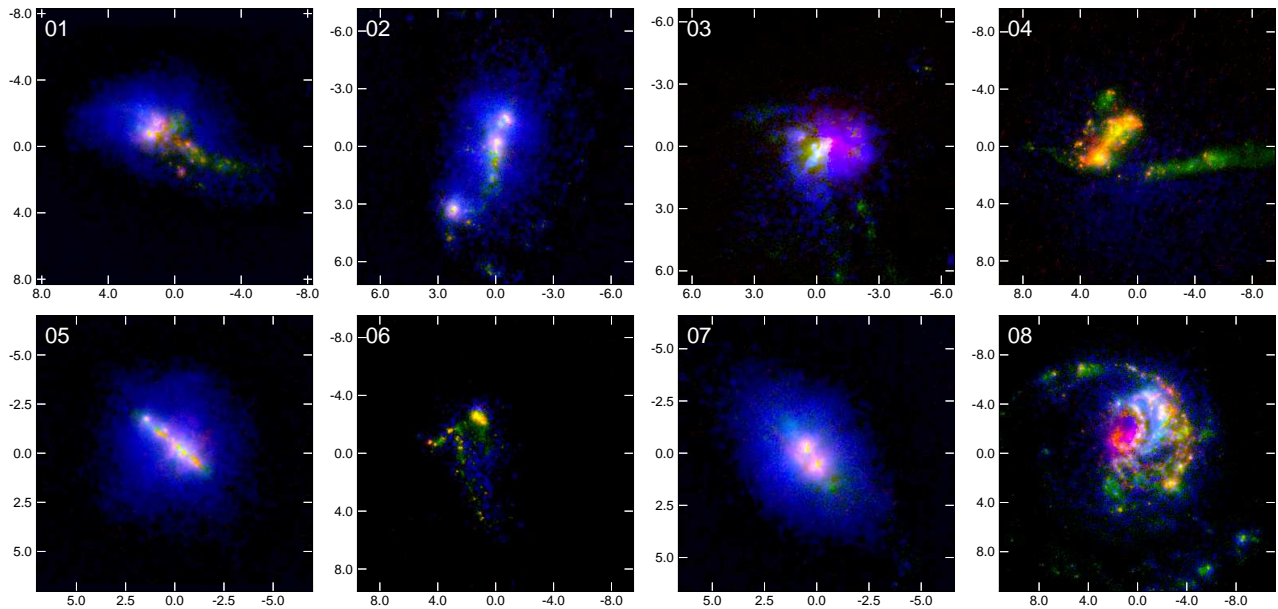}
\caption{False-color images of the LARS galaxies 01 to 08. Red encodes 
continuum-subtracted \halpha, green the FUV continuum, and blue shows 
continuum-subtracted \lya. Images have been 
adaptively filtered to show detail. Scales in kpc are given on the 
side. Intensity scales are logarithmic, with intensity cut levels set 
to show detail.  }
\label{fig:rgb1}
\end{figure*}

Because \hi\ is often found at distances that
exceed the size of stellar disks and star-forming regions 
\citep{Yun1994,Meurer1996,Cannon2004}, characteristic \lya\ scale 
lengths may be expected to be substantially larger than those of,
for example, the FUV continuum or \halpha.
Indeed this has been well observed
at high $z$ (e.g. \citealt{Fynbo2001,Rauch2008,Steidel2011}, although 
see also \citealt{Feldmeier2013})	and low 
\citep{Mas-Hesse2003,Ostlin2009}, and studied extensively by simulation 
\citep{Laursen2009b,Barnes2010,Zheng2011,Verhamme2012}. 

In this \emph{Letter} we present images from the \emph{Lyman alpha 
Reference Sample} (LARS). The LARS program (\"Ostlin et al., in prep;
Hayes et al., in prep) is targeting 14 UV-selected star-forming 
galaxies in the nearby Universe, all of 
which have been imaged in \lya, \halpha, \hbeta, and five UV/optical 
continuum bands. Many other observations, both in hand and ongoing, 
are providing gas covering fractions and kinematics, and measuring the 
\hi\ mass and extent directly. HST imaging allows us to probe spatial 
scales down to 
28~pc in individual galaxies, quantify the extent of \lya, and
compare it with other wavelengths and derived properties. This 
\emph{Letter} discusses the extension of \lya\ radiation.
In Section~\ref{sect:data} we briefly summarize the data and show the 
new images. In Section~\ref{sect:extent} we quantify the sizes of the 
galaxies in \lya, FUV, and \halpha, and discuss them with reference to 
high-$z$ measurements in Section~\ref{sect:highz}.
In Section~\ref{sect:corr} we show how a low 
dust content seems to be a necessary prerequisite in order to produce
this extended emission. We assume a cosmology of 
$(H_0, \Omega_\mathrm{M}, \Omega_\Lambda) = 
(70~\mathrm{km~s}^{-1}~\mathrm{Mpc}^{-1}, 0.3, 0.7)$.

\section{LARS Images}\label{sect:data}

LARS consists of 14 star-forming galaxies selected by FUV luminosity from
the GALEX all-sky surveys, and imaged with Hubble Space Telescope 
cameras ACS/SBC, ACS/WFC, and WFC3/UVIS. The 
sample selection, observations, and data processing 
are described in detail in \"Ostlin et al. (in prep). 
FUV luminosities range between 
$\log (L_\mathrm{FUV}/L_\odot) = 9.2$ and 10.7, overlapping much of the
luminosity range of Lyman-break Galaxy (LBG) surveys, and are listed
in Table~\ref{tab:quants}.

We use the \emph{Lyman alpha
eXtraction software} ({\tt LaXs}, \citealt{Hayes2009}) to produce 
continuum-subtracted \lya\ and \halpha\ images, corrected for 
underlying stellar absorption and contamination from \nII. 
In 1 arcsec square boxes away from the targets we 
measure r.m.s. background noise of 
$5.7 \times 10^{-19}$~\ergseccm\ in \lya,
$2.1 \times 10^{-21}$~\ergseccmaa\ in the FUV, and 
$6.8 \times 10^{-19}$~\ergseccm\ in \halpha. 
Total \lya\ luminosities range 
from 0 (non-detection) and $2 \times 10^{43}$~\ergsec\
with a median of $8.1\times 10^{41}$~\ergsec; roughly seven of the 
objects would be recovered by the deepest \lya\ surveys (Hayes et al. 
in prep).

\begin{figure*}
\centering
\includegraphics[scale=0.42]{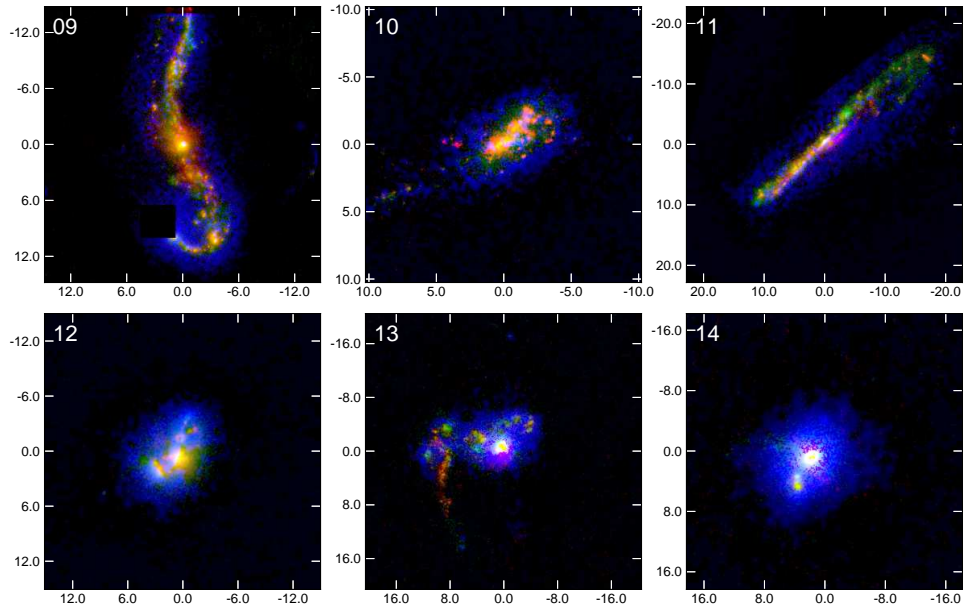}
\caption{Same as Figure~\ref{fig:rgb1} except for LARS galaxies 09 to 14.
The black square in LARS\,09 masks a UV-bright field star. }
\label{fig:rgb2}
\end{figure*}

We present our first imaging results in this paper as a series of RGB 
composite images in Figures~\ref{fig:rgb1} and \ref{fig:rgb2}. In 
green we encode the far UV
continuum, which traces the unobscured massive stars, and roughly 
incorporates the sites that produce the ionizing photons. In the 
red we show continuum-subtracted \halpha, which traces the 
nebulae where the aforementioned ionizing photons are 
reprocessed into the recombination line spectrum. 
The continuum subtracted \lya\ observation is encoded in blue. The 
images have been adaptively smoothed using a variable Gaussian kernel 
(\texttt{FILTER/ADAPTIVE} in \texttt{ESO/MIDAS}), 
in order to enhance positive regions of low surface brightness emission. 
The intensity scaling of all the images is logarithmic, and the 
levels are set to show the maximum of structure and the level at which 
the faintest features fade into the background. 

Immediately it can be seen that \lya\ morphologies bear limited 
resemblance to those of the FUV and \halpha. In some cases 
\lya\ appears to be almost completely absent: LARS\,04 and 06 in 
particular show only small hints of \lya\ emission that contribute 
negligibly towards filling in the global absorption, and the 
composites are dominated by UV and \halpha\ light. \lya\ is 
strongly absorbed, particularly in the central regions of these 
objects. Others show copious \lya\ emission and reveal morphological 
structures that are not seen at other wavelengths. Most obviously, 
LARS\,01, 02, 05, 07, 12, and 14 show large-scale halos of \lya\ 
emission that completely encompass the star-forming regions, although
the same phenomenon is visible to some extent in all the objects,
even the absorbers. 

We have discussed this extended \lya\ emission in depth in the past
\citep{Hayes2005,Hayes2007,Atek2008,Ostlin2009}. However now, with an 
observational setup that is more sensitive to faint levels of \lya\ 
and a larger and UV-selected sample (\"Ostlin et al, in prep), we
are able to robustly quantify and contrast these sizes and the relative
extension of \lya. 

\section{Apertures, Sizes and Global Quantities}\label{sect:extent}

In order to quantify the sizes of the galaxies at various 
wavelengths, we adopt the Petrosian radius \citep{Petrosian1976}
with index of $\eta=0.2$: i.e. the radius, $R$, at which the local 
surface brightness is 20 percent the average surface 
brightness inside of $R$. In Hayes et al. (in prep) we will show the 
\lya\ extent of some objects to be so large that ACS/SBC cannot
capture the full flux, and hence measurements like 50 percent light 
radius are not robust. 
Indeed Petrosian radii were developed to be 
depth-independent measures of size.
We note from experimentation, however, that very
similar conclusions are reached using other definitions. 
The choice of 
$\eta=0.2$ gives a size for every \lya-emitting galaxy in the 
sample except LARS\,09, for which even at the full extent of the SBC 
we do not come close to crossing the $\eta=0.2$ threshold. We reach the 
edge of the detector at 
$\eta \sim 1$ ($R>12$~kpc) and can expect the true extent of \lya\
				to be 
much larger. For the 11 galaxies in which \rptlya\ is well 
measured, its determination is robust, and would not change  
were the observations deeper or the field-of-view larger.  
\rpt, is computed for \lya, \halpha, and the FUV continuum, and listed 
in Table~\ref{tab:quants}. 
Based upon
aperture-matched \halpha\ and \hbeta\ imaging and standard Case B 
assumptions, we recover up to 60~\% of the intrinsic \lya\
flux, although the median value is just $\sim 3$~\% (Hayes et al. in prep).

\begin{deluxetable*}{clllccccccccc}
\tabletypesize{\scriptsize}
%\rotate
\tablecaption{The LARS sample: properties and sizes.\label{tab:quants}}
\tablehead{
\colhead{LARS} & 
		\colhead{Common name} & 
		\colhead{R.A.} & 
		\colhead{Dec.} & 
		\colhead{$z$} & 
		\colhead{\lfuv} & 
		\colhead{\rptfuv} & 
		\colhead{\rptha} & 
		\colhead{\rptlya} & 
		\colhead{$R_{\mathrm{chip}}^{\mathrm{SBC}}(z)$} &
		\colhead{\rext} & 
		\colhead{$\beta$-slope} & 
		\colhead{\halpha/\hbeta} \\
\colhead{ID} & 
		\colhead{} & 
		\colhead{$h$:$m$:$s$} & 
		\colhead{$d$:$m$:$s$} & 
		\colhead{} & 
		\colhead{\lsun} & 
		\colhead{kpc} & 
		\colhead{kpc} & 
		\colhead{kpc} & 
		\colhead{kpc} & 
		\colhead{} & 
		\colhead{} & 
		\colhead{}  \\
\colhead{(1)} & 
		\colhead{(2)} & 
		\colhead{(3)} & 
		\colhead{(4)} & 
		\colhead{(5)} & 
		\colhead{(6)} &  
		\colhead{(7)} &  
		\colhead{(8)} & 
		\colhead{(9)} &  
		\colhead{(10)} & 
		\colhead{(11)} &  
		\colhead{(12)} &  
		\colhead{(13)}}  
\startdata
01 & Mrk\,259	       & 13:28:44.0 & +43:55:49.9 & 0.028 & 9.92 & 1.18 & 1.29 & ~~~4.36    & 7.87 & ~~3.37    & --1.83 & 3.08 \\ 
02 & \nodata         & 09:07:04.9 & +53:26:56.5 & 0.030 & 9.48 & 1.12 & 1.17 & ~~~2.67    & 8.41 & ~~2.27    & --2.02 & 3.08 \\ 
03 & Arp\,238	       & 13:15:35.1 & +62:07:27.2 & 0.031 & 9.52 & 0.84 & 0.97 & ~~~0.75    & 8.68 & ~~0.77    & --0.57 & 5.18 \\ 
04 & \nodata         & 13:07:28.2 & +54:26:50.7 & 0.033 & 9.93 & 3.79 & 1.57 & ~~~\nodata & 9.22 & ~~\nodata & --1.76 & 3.48 \\ 
05 & Mrk\,1486	     & 13:59:51.0 & +57:26:23.0 & 0.034 & 10.0 & 0.93 & 1.24 & ~~~3.24    & 9.49 & ~~2.61    & --2.09 & 3.06 \\ 
06 & KISSR\,2019	   & 15:45:44.5 & +44:15:49.9 & 0.034 & 9.20 & 3.65 & 0.66 & ~~~\nodata & 9.48 & ~~\nodata & --1.85 & 2.96 \\ 
07 & IRAS\,1313+2938 & 13:16:03.9 & +29:22:54.2 & 0.038 & 9.75 & 0.85 & 0.89 & ~~~3.01    & 10.5 & ~~3.37    & --1.94 & 3.37 \\ 
08 & \nodata         & 12:50:13.7 & +07:34:44.2 & 0.038 & 10.2 & 5.01 & 3.89 & ~~~4.35    & 10.5 & ~~1.12    & --0.90 & 4.09 \\ 
09 & IRAS\,0820+2816 & 08:23:54.9 & +28:06:22.8 & 0.047 & 10.5 & 5.00 & 4.21 & $>$12.0    & 12.9 & $>$2.85 & --1.52 & 3.48 \\ 
10 & Mrk\,0061	     & 13:01:41.5 & +29:22:53.2 & 0.057 & 9.74 & 2.34 & 2.63 & ~~~5.49    & 15.5 & ~~~2.08    & --1.36 & 3.93 \\ 
11 & \nodata         & 14:03:47.1 & +06:28:15.0 & 0.084 & 10.7 & 8.00 & 6.81 & ~~~15.5    & 22.1 & ~~~2.27    & --1.50 & 4.60 \\ 
12 & SBS\,0934+547   & 09:38:13.5 & +54:28:25.3 & 0.102 & 10.5 & 1.78 & 2.03 & ~~~7.06    & 26.3 & ~~~3.48    & --1.92 & 3.21 \\ 
13 & IRAS\,0147+1254 & 01:50:28.4 & +13:08:59.2 & 0.147 & 10.6 & 3.83 & 4.68 & ~~~8.12    & 36.0 & ~~~1.74    & --1.53 & 4.07 \\ 
14 & \nodata         & 09:26:00.3 & +44:27:36.0 & 0.181 & 10.7 & 0.79 & 1.62 & ~~~5.86    & 42.7 & ~~~3.62    & --2.22 & 3.13 
\enddata
\tablecomments{Coordinates (3 and 4) are J2000. Redshifts (5) are 
derived from SDSS. (6) are $\log(\nu L_\nu)$ in solar luminosities. 
(7--9) are the Petrosian radii with $\eta=0.2$, \rpt. 
(10) shows the physical scale corresponding to an angular
size of 14 arcsec at the redshift of each galaxy -- this corresponds to
half the diametric size of the ACS/SBC (28 arcsec / 2) and describes 
the maximum usable scale to which we can probe \lya. (11) gives the relative 
extension of \lya\ relative to\halpha. 
(12) gives the UV slope, $\beta$, derived from HST imaging. 
(13) gives the \halpha/\hbeta\ ratio, derived from SDSS spectroscopy.  }
\end{deluxetable*}

\begin{figure*}
\centering 
\includegraphics[scale=0.7, clip=true, trim=0mm 0mm 0mm 0mm]{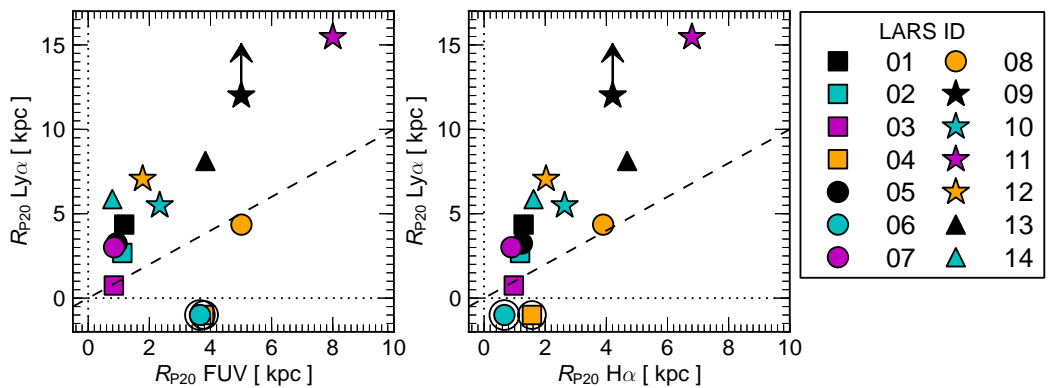}
\caption{Comparison of the Petrosian radii ($\eta = 0.2$), \rpt,  in 
continuum-subtracted \lya, \halpha, and the UV continuum. In cases 
where a galaxy is a net \lya\ absorber its size is undefined, and 
\rptlya\ has been set to a small negative value -- it could in 
principle also be very large. The 
\emph{Left}  panel shows how \lya\ sizes compare with the FUV, which 
can be similarly derived at high-$z$. The \emph{Right} panel 
makes the same comparison against \rptha, which directly contrasts 
the intrinsic and emitted \lya\ sizes. When net \lya\ emission is 
found it is systematically extended, taking mean values of 
\rptlya/\rptfuv=2.9 (left) and \rptlya/\rptha=2.4 (right).}
\label{fig:sizes}
\end{figure*}

We compare the light radii graphically in Figure~\ref{fig:sizes}. 
The plots show \rptlya\ vs. \rptfuv, a comparison that could be 
made at high-$z$, and \rptlya\ vs. \rptha, a comparison that more 
directly  conveys the difference between the observed and intrinsic 
\lya\ sizes. Clearly, though, there is little difference in the 
result: \lya\ radii are, on average, substantially larger than 
corresponding FUV or \halpha\ radii. In Table~\ref{tab:quants} we also
report the \emph{Relative Petrosian Extension} of \lya\ compared to 
\halpha, \rext, which is simply defined as \rptlya /\rptha. 12 galaxies
show net emission of \lya, where all except for one (LARS\,03) has
\rext$>1$. 
The galaxy with the largest extension is LARS\,14, for 
which we measure \rext=3.6.
It is not clear whether globally absorbing 
galaxies LARS\,04 and 06 become emitters on larger scales, but if so 
their \rptlya\ must be larger than the radius of the SBC chip, implying
that \rext\ must exceed 5.3 and 13.4, respectively. That would make 
them the most extended objects in the sample. 
Excluding these two galaxies, and also LARS\,09 for which we can only 
provide a lower limit, the sample mean (median) is computed as 2.43 (2.28). 

\section{Relevance for high-redshift studies }\label{sect:highz}

It is important to note that FUV radii imply that all the galaxies 
would be effectively unresolved by ground-based observations 
if they were at $z\gtrsim 2$. The largest is 8~kpc, which corresponds
to the 1 arcsec resolution that could be expected from the seeing. 
However one of the objects has a \lya\ radius of 15.5~kpc: recovering 
this total flux at $z\sim 2$ would require an aperture of at least 2 
arcsec. Some objects are also highly elongated and were they 
pushed to the high-$z$ Universe, much of their \lya\ could also
be unmeasured if circular apertures are used. 

\lya\ emission more extended than the FUV has been 
reported in numerous high-$z$ samples. \citet{Fynbo2003} 
remarked upon a few such objects at the brighter end of the luminosity 
distribution of the 27 narrowband selected galaxies, and the 
extremely deep spectroscopic observations of \citet{Rauch2008} 
uncovered 28 \lya\ galaxies, ten of which were classified as extended.
Samples of \lya\ blobs 
\citep[e.g.][]{Matsuda2012,Prescott2012} may be many times the size 
of their counterpart galaxies, if indeed counterparts are 
identified at all. 
Here we report that every galaxy in the sample that emits \lya\ does so
by producing a halo; on average the halo is 
over twice the linear size of \halpha\ and the FUV.

By stacking narrowband images of LBGs at 
$\langle z\rangle=2.65$, \citet{Steidel2011} reported \lya\ halos that
extend many tens of kpc, probably probing the neutral circumgalactic 
medium (CGM) out to the virial radius. Subdividing the full sample by 
\lya\ properties, the halos at radii larger than 20--30 physical kpc 
show very similar scale lengths in all subsamples (although different
central surface brightnesses), even when central \lya\ absorption is 
found. At small radii the subsamples exhibit profiles
that differ markedly, dropping rapidly to $\sim 0$ for the 
\lya-absorbing sample but steepening by varying degrees in all 
others. Even the steepest central profiles, however, still run 
much flatter than those of the stellar continuum, and this 
change likely marks the onset of higher density gaseous disks or 
similar.

From the various $z\approx 2.7$ \lya\ profiles of 
\citet{Steidel2011} we calculate \rptlya\ using the same method as for
our sample, and dividing by \rptfuv\ from the continuum profile
we obtain \rext\ (now relative to the UV). These raw values range 
between \rext=3.8 for the non-LAEs, and 5.9 for the LAE-only sample,
and are notably bigger than our largest \rext. 
However, under the assumption that the inner and outer profiles mark 
physically different regimes that may not be the same in low-$z$ 
galaxies, we also subtract the exponential halo fits of \citet{Steidel2011}
and repeat the exercise; this yields a range of \rext=0.84 to 2.0. This is
now smaller than many of our values, although close to the average and the 
dispersion of the high-$z$ sample is obviously lost in the stacking 
process. On the other hand, 
the UV continuum profile of \citet{Steidel2011} is 
dominated by atmospheric seeing. If we instead use the continuum 
effective radius of BM/BX galaxies and LBGs from HST imaging 
\citep{Mosleh2011} we compute \rptfuv$\approx 5$~kpc, which would increase all 
the \rext\ quoted above by a factor of 2.5. \rext\ from the raw data
would then become much larger than we measure in the local universe (up
to 15), and \rext\ in halo-subtracted profiles that are roughly 
consistent (2.1 to 4.8).

\begin{figure*}[t!]
\centering 
\includegraphics[scale=0.6, clip=true, trim=0mm 0mm 0mm 0mm]{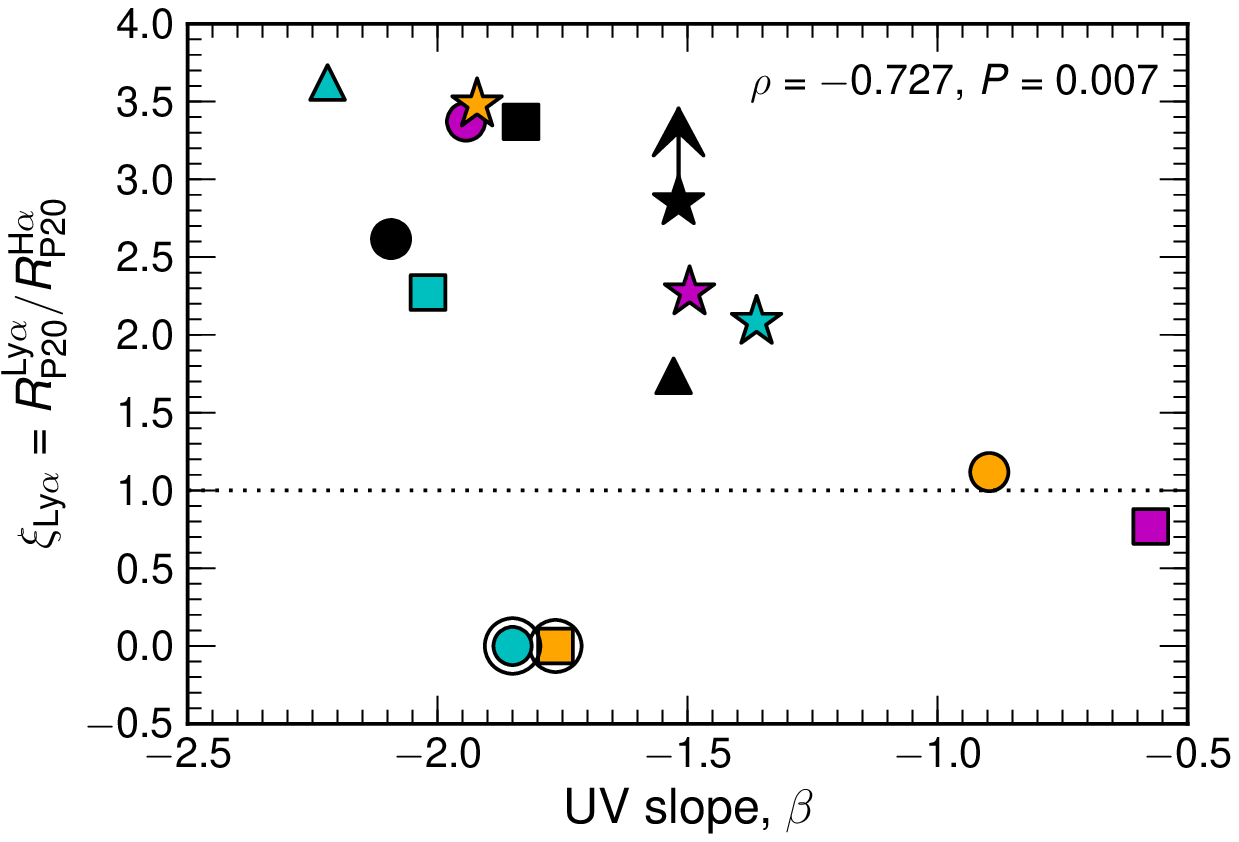}
\includegraphics[scale=0.6, clip=true, trim=0mm 0mm 0mm 0mm]{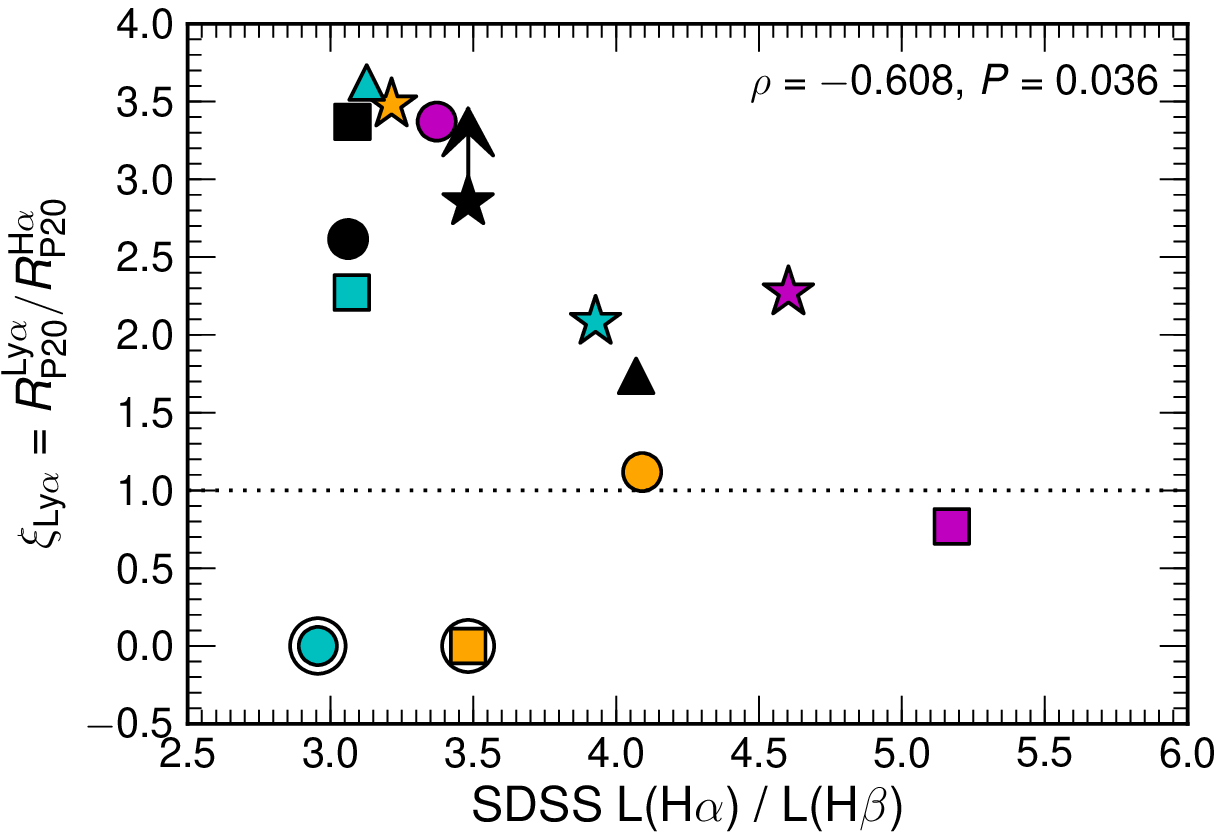}
\caption{Correlations between the \lya\ vs. \halpha\ extension, \rext, 
and measures of the galaxy reddening. The \emph{Left} panel shows the 
UV slope, $\beta$ measured from HST imaging, while the \emph{Right} 
panel shows the nebular attenuation measured from SDSS 
spectroscopy at \halpha\ and \hbeta. Net absorbing galaxies are set to
zero and ringed, but could in principle also be very extended. 
Rank correlation coefficients of the Spearman $\rho$ test and the 
associated probability of the no-correlation
hypothesis (accounting for ties and the small sample) are given in the 
top-right corners. Symbols are the same as Figure~\ref{fig:sizes}. }
\label{fig:corr}
\end{figure*}

LARS observations probe scales far below the tens of kpc sampled 
at high-$z$ on a case-by-case basis.
The galaxies likely include the range between, or roughly bracketing, 
the averaged subsamples of \citet{Steidel2011}. Our imaging also 
suggests this extension to be a very common property of 
\lya-emitting galaxies, and its onset begins almost immediately in the
inner few kpc: we find seven galaxies with FUV Petrosian radii below 
2~kpc, five of which have corresponding \lya\ radii three times larger. 

It is also noteworthy that \citet{Steidel2011} 
find different median dust attenuations for the \lya-emitting and 
non-emitting subsamples, almost precisely as we did in 
\citet{Hayes2010}. LAEs, which show extended central peaks, were determined
to have stellar \ebv=0.09 magnitudes (c.f. 0.085 in 
\citealt{Hayes2010}) while absorbers show \ebv=0.19 (c.f. 0.23 for our 
\halpha-selected sample). Adopting the prescription of 
\citet{Meurer1999} the stellar \ebv\ measurements for the Steidel et al. 
samples correspond to $\beta$
slopes\footnote{UV continuum flux density, parameterized by a power-law of
the form $f_\lambda \propto \lambda^\beta$.} of $-1.77$ (LAEs) and 
$-1.27$ (\lya\ absorbers). Bluntly accounting for a factor of 2.27 that
connects stellar \ebv\ to its nebular equivalent 
in local starbursts \citep{Calzetti2000}, the same stellar \ebv\ would 
equate to \halpha/\hbeta\ ratios of 3.5 (LAEs) and 4.4 (absorbers). 
In the next Section we will show case-by-case that \lya\ halos 
systematically become more extended with decreasing dust contents.

\section{Lyman alpha extension and dust contents}\label{sect:corr}

In Hayes et al. (in prep) we compute many global properties 
for the sample, in order to study the processes complicit
in \lya\ transport. Indeed that paper will 
include a complete analysis of correlations between \lya\ transmission,
halo sizes, and many other properties; in this 
\emph{Letter} we restrict ourselves to observables that scale with the 
dust content. It is noteworthy for the moment, however, that we find 
no correlation between \rext\ and the total \lya\ luminosity. 
In Figure~\ref{fig:corr} we show how \rext\ 
compares with both the UV continuum slope $\beta$ 
and the \halpha/\hbeta\ ratio. We note that the SDSS fibers are on 
average smaller than the \lya\ radii, but do capture 
the bulk of the nebular emission, and fluxes can easily be measured 
without contamination of \nII\ and stellar absorption. 
Since \citet{Meurer1999} $\beta$ has been used almost ubiquitously as a
proxy of stellar attenuation in high-$z$ galaxies; here we measure
$\beta$ from aperture-matched HST imaging using the FUV 
(SBC/F140LP or F150LP) and the $U-$band (UVIS/F336W or F390W) filters.
With colors between $\beta\approx -2.2$ and $-0.6$ our 
objects have similar UV slopes to the vast majority of those found in 
$z=2-4$ \lya-emitting galaxies \citep{Blanc2011}. Similarly the 
\halpha/\hbeta\ ratio is the canonical probe of nebular reddening
(i.e. that which is to zeroth order expected for \lya) used in studies
of low-$z$ and Galactic nebulae. $\beta$ and \halpha/\hbeta\ are listed in 
Table~\ref{tab:quants}. 

Both measures of dust content strongly anti-correlate with \rext\ 
although the sample is small ($N=12$ defined sizes in \lya). To assess 
its significance we compute the Spearman rank correlation 
coefficient, $\rho$, which yields $\rho = -0.73$ and  $-0.61$ for the 
anti-correlation of \rext\ with $\beta$ and \halpha/\hbeta, 
respectively. This corresponds to likelihoods of the null hypothesis 
-- that this correlation arises purely by chance -- amounting to 0.7 
percent (UV slope), and 3.6 percent (\halpha/\hbeta).

The halo--dust phenomenon appears not to be a direct effect of 
radiative transfer. We have performed new test simulations with the 
\texttt{McLya} code \citep{Verhamme2006}, by tuning the gas-to-dust
ratio in the synthetic galaxy of \citet{Verhamme2012}. 
Indeed the surface brightness does scale with dust abundance but the 
light profile (therefore \rptlya) does not, and the
\rext--dust trend must be a secondary correlation. A scenario
is needed in which galaxies decrease the relative size of their \hi\ 
envelopes as the absolute dust content increases. A sequence in
which neutral gas settles into the galaxy (reducing \rext) and 
subsequently forms stars (creating more dust) would explain the 
trend, but without yet having obtained spatially resolved \hi\ data 
this is conjecture.

Scattering also has the potential to spread \lya\ over such an area 
that its surface brightness decreases greatly. In such a case, 
scattered radiation measured at large radii may not be sufficient to 
recover flux from a broad central absorption, making \rext\ 
observationally undefined when it is actually very large. The trend 
of \rext\ increasing in bluer galaxies, then, is 
also able to explain the undefined sizes of LARS 04 and 06, 
at their measured dust abundance. Similar considerations would also
explain the non-detection of \lya\ in local gas-rich but  metal- and 
dust-poor dwarf starbursts such as 
{\sc i}\,Zw\,18 and SBS\,0335--052 
\citep{Kunth1994,Mas-Hesse2003,Ostlin2009}, as discussed 
in \citet{Atek2009izw18}.

We have empirically shown before 
\citep{Atek2009fesc,Hayes2010} that the global escape fraction of \lya\
photons anti-correlates strongly with attenuation (also
\citealt{Kornei2010} in LBGs). We now demonstrate 
that at lower \ebv, the more strongly emitting galaxies 
are likely to also spread their \lya\ over larger surfaces. 
Thus while they do transmit more of their \lya, it may be that 
more of the transferred \lya\ is observationally lost 
outside photometric apertures. This may also explain the 
lack of correlation between \lya/\hbeta\ and \ebv\ observed by 
\citet{Giavalisco1996}, compared to trends seen in other samples: the 
aperture of the IUE probed just 
3~kpc at $z=0.01$ and if more \lya\ is lost in bluer galaxies 
the \lya/Balmer ratios would be artificially lowered in in such systems.
This could in part mask an underlying correlation. 
By a similar token, galaxies that can very efficiently scatter 
\lya\ photons may not be recovered at all, despite frequently showing 
very blue UV colors. Determining precisely how \lya\ profiles are 
modified for a given set of host properties will provide a cornerstone
for interpreting future large high-$z$ surveys.

\acknowledgments
M.H. received support from Agence Nationale de la recherche
bearing the reference ANR-09-BLAN-0234-01.
G.\"O. is a Swedish Royal Academy of Sciences research fellow supported
by a grant from Knut and Alice Wallenberg foundation, and also 
acknowledges support from the Swedish research council (VR) and the 
Swedish National Space Board (SNSB).
A.V. benefits from the fellowship `Boursi\`ere d'excellence de 
l'Universit\'e de Gen\`eve'. 
H.A. and D.K. are supported by the Centre National d'\'Etudes Spatiales (CNES) 
and the Programme National de Cosmologie et Galaxies (PNCG).
I.O. acknowledges the Sciex fellowship.
H.O.F. acknowledges financial support from CONACYT grant 129204, and Spanish FPI grant BES-2006-13489.
H.O.F. and J.M.M.H. are partially funded by Spanish
MICINN grants CSD2006-00070 (CONSOLIDER GTC), AYA2010-21887-C04-
02 (ESTALLIDOS) and AYA2011-24780/ESP.
We thank C. Steidel for making the 
high-$z$ \lya\ profiles available for our comparisons in 
Section~\ref{sect:highz}.

{\it Facilities:} \facility{HST (ACS,WFC3)}.

\end{document}